\documentclass[prl,twocolumn,superscriptaddress,preprintnumbers,floatfix,nofootinbib]{revtex4-2}
\usepackage[T1]{fontenc}
\usepackage{epsfig}
\usepackage{graphicx}
\usepackage[english]{babel}
\usepackage{hyphenat}
\usepackage{amsmath}
\usepackage{amssymb}
\usepackage{mathtools}
\usepackage{mathrsfs}
\usepackage{slashed}
\usepackage{epstopdf}
\usepackage{xcolor}
\usepackage{braket}
\usepackage{booktabs}
\usepackage[normalem]{ulem}
\definecolor{lcolor}{rgb}{0.,0.0,0.}
\definecolor{citcolor}{rgb}{0,0.,0.5}
\usepackage[breaklinks,colorlinks,urlcolor=blue,citecolor=blue,linkcolor=blue]{hyperref}
\usepackage{multirow}
\usepackage{ltablex}
\usepackage{soul}

\usepackage[e]{esvect}

\newcommand{\secn}[1]{Section~1}
\newcommand{\appn}[1]{Appendix~1}

\long\def\comment#1{ }

\def\and{\quad\text{and}\quad}

\def\0{{\boldsymbol 0}}
\def\1{{\boldsymbol 1}}

\def\0{{\boldsymbol 0}}

\newcommand{\mcal}{\mathcal}

\newcommand\n{\mathbf{n}}

\renewcommand{\part}{{\rm part}}

\newcommand{\be}{\begin{equation}}
\newcommand{\ee}{\end{equation}}
\newcommand{\bes}{\begin{subequations}}
\newcommand{\ees}{\end{subequations}}
\newcommand{\bea}{\begin{eqnarray}}
\newcommand{\eea}{\end{eqnarray}}

\def\bea#1\eea{\begin{align}#1\end{align}}
\newcommand{\bef}{\begin{figure}[h!tb]\centering}
\newcommand{\eef}{\end{figure}}

\allowdisplaybreaks

\begin{document}

\title{Exploring QCD Matter through Energy Correlators of Soft Hadrons}
\author{Yiyang Peng}
\email{yiyangpeng@stu.pku.edu.cn}
\affiliation{School of Physics, Peking University, Beijing 100871, China}

\author{Jo\~{a}o Barata}
\email{jlourenco@bnl.gov}
\affiliation{Physics Department, Brookhaven National Laboratory, Upton, NY 11973, USA}

\author{Andrey V. Sadofyev}
\email{andrey.sadofyev@ehu.eus}
\affiliation{Department of Physics, University of the Basque Country UPV/EHU, P.O. Box 644, 48080 Bilbao, Spain}
\affiliation{EHU Quantum Center, University of the Basque Country UPV/EHU, P.O. Box 644, 48080 Bilbao, Spain}
\affiliation{IKERBASQUE, Basque Foundation for Science, Plaza Euskadi 5, 48009 Bilbao, Spain}

\author{Huichao Song}
\email{huichaosong@pku.edu.cn}
\affiliation{School of Physics, Peking University, Beijing 100871, China}
\affiliation{Center for High Energy Physics, Peking University, Beijing 100871, China}

\begin{abstract}
We propose energy-energy correlators (EECs) of soft hadrons and their azimuthal harmonics as new soft observables to probe the collective flow and correlations beyond flow for large and small systems created in high energy nuclear collisions. We separate the soft EEC into a disconnected contribution determined by the mean energy flow and a connected contribution encoding correlations beyond it.  Viscous hydrodynamic simulations demonstrate that the soft EEC of a fluid system is governed by the disconnected contribution, while its elliptic and triangular harmonics respond to the initial geometry and shear viscosity as traditional flow harmonics do. Comparisons between iEBE-VISHNU and PYTHIA for p+p, O+O, and Pb+Pb collisions show that, for both the EEC and its elliptic harmonics, the disconnected contribution dominates at large angles in all systems, whereas connected correlations beyond the mean energy flow grow toward small angles and small systems. Soft EECs thus provide a unique window into collective flow and correlations beyond it across collision systems.
\end{abstract}

\maketitle

\noindent\textit{\textbf{Introduction:}} 
Collective flow is an important signal for the quark-gluon plasma (QGP) formation in high energy nuclear collisions. The initially far-from-equilibrium deconfined matter approaches local equilibrium on microscopic time scales and develops strong collective expansion driven by pressure gradients of the QGP. This evolution is well described by relativistic hydrodynamics, which has achieved remarkable success in describing a wide range of flow observables~\cite{Kolb:2003dz,Teaney:2009qa,Schenke:2010rr,Gale:2012rq,Heinz:2013th,Hirano:2012kj,
Florkowski:2017olj,Song:2017wtw,Shen:2020mgh,Heinz:2024jwu}. This makes hydrodynamics and flow  key tools for constraining the transport properties of QGP and the initial-state geometry and fluctuations in large collision systems such as Pb+Pb. At the same time, understanding how the picture of collective flow evolves toward smaller collision systems such as O+O and p+p collisions, where correlations and non-flow effects become increasingly important, remains an active question~\cite{Grosse-Oetringhaus:2024bwr}.

Traditional flow observables are constructed from the final hadron distributions in momentum space, which reflect the macroscopic expansion of the evolving system but typically suppress localized microscopic correlations. Energy correlators, however, admit a direct field-theoretic definition in terms of asymptotic energy fluxes and retain energy-weighted correlations over a broad angular range~\cite{Basham:1978zq,Basham:1979gh,Basham:1977iq,Basham:1978bw}, see also~\cite{Moult:2025nhu} for a recent review. Varying the angular separation then gives access to dynamics ranging from microscopic correlations at small angles to the large-scale structure of collective flow at large angles. They thus provide a natural extension of traditional flow analysis while preserving sensitivity to correlations beyond collective flow~\cite{Barata:2026pgh,Barata:2026aak}.

\begin{figure}[h!]
    \centering
    \includegraphics[width=0.6\columnwidth]{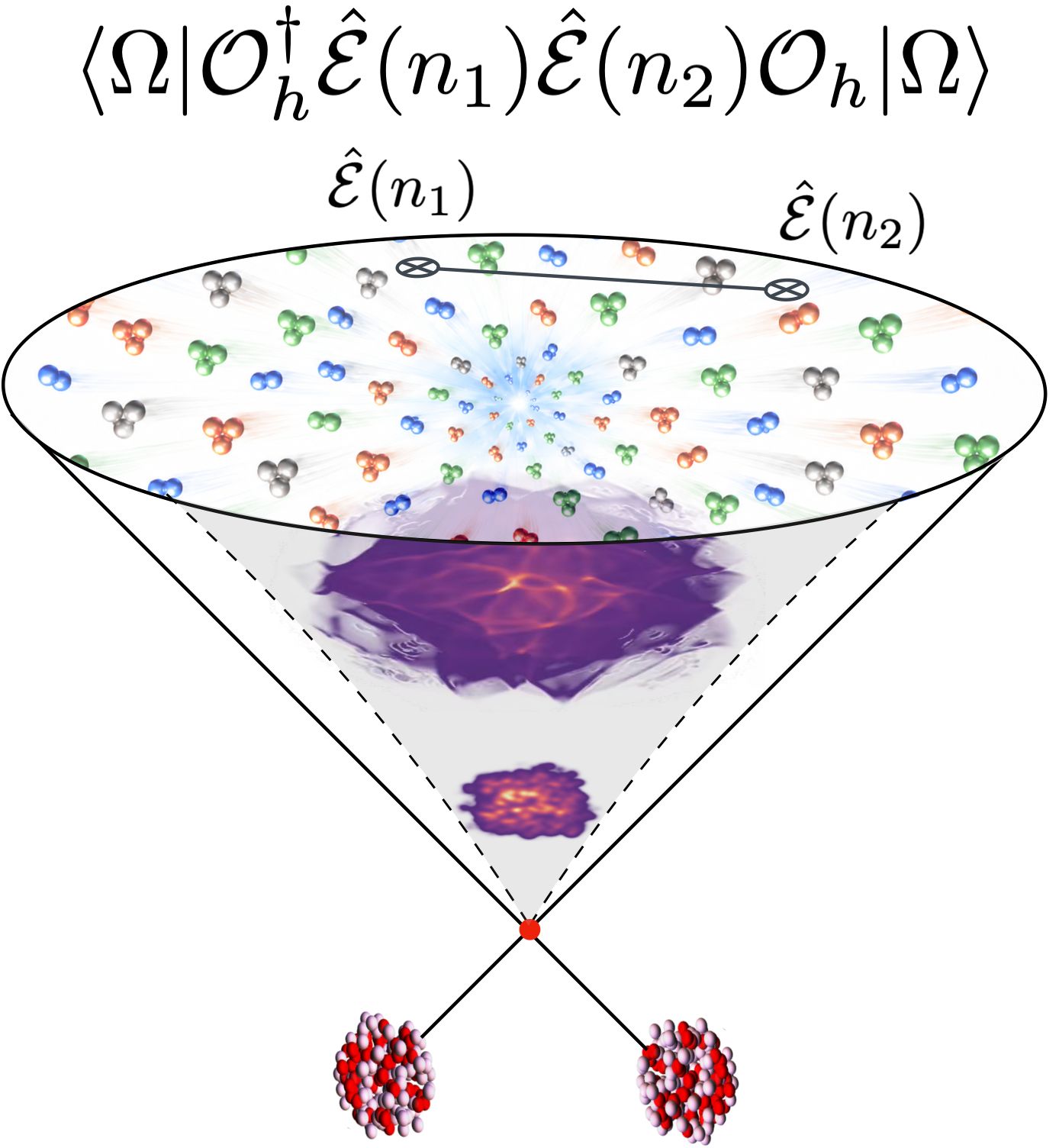}
    \caption{Space-time evolution of the bulk matter produced in high energy nuclear collisions. Correlations of energy flow operators ${\hat{\mathcal{E}}}(\mathbf{n})$ are measured on the hadrons produced by the hot QCD matter, generated by a many-body operator acting on the vacuum, i.e., $\hat{\mathcal{O}}_h |\Omega \rangle$.}
    \label{fig:cartoon}
\end{figure}

Formally, energy correlators are built from the energy flow operator, defined as \cite{Ore:1979ry,Sveshnikov:1995vi,Tkachov:1995kk,Korchemsky:1999kt}
\begin{align}\label{eq:def_Ec}
{\hat{\mathcal{E}}}(\mathbf{n}) = \lim_{r\to\infty} r^2 \int_0^\infty dt\, n^i {\hat{T}}^{0i}(t,r\mathbf{n})\, ,
\end{align}
where $\mathbf{n}$ is a unit three-vector labeling a point on the sphere at large $r$, see Fig.~\ref{fig:cartoon}. Their definition retains a direct connection to QCD while providing a natural realization in terms of calorimetric measurements. More generally, studies of energy-flow correlators in complex many-body states have shown that their angular structure is constrained by general theoretical considerations, see e.g.
~\cite{Hofman:2008ar,Delacretaz:2018cfk,Chicherin:2023gxt,Cuomo:2025pjp,
Barata:2024ukm,Barata:2025fzd,Barata:2026pgh,Riembau:2024tom,Georgoudis:2026vjm,Grassi:2026tdb}. 
 
In this \textit{Letter}, we bring this perspective to high energy nuclear collisions and propose the energy--energy correlator (EEC) of soft hadrons and its azimuthal harmonics as new observables to probe collective flow and correlations beyond flow for large and small collision systems. We separate the soft EEC into a disconnected part determined by the mean energy flow and a connected part encoding correlations beyond it. For fluid systems well described by hydrodynamics, the EEC and its harmonics are dominated by the disconnected contribution, and EEC harmonics retain sensitivity to the initial-state geometry and the shear viscosity of the medium as traditional flow harmonics do. As the system becomes smaller and non-flow correlations grow increasingly important, the connected contribution becomes progressively more significant at small angular separations. The soft EEC and its azimuthal harmonics therefore provide a unique way to probe collective flow and correlations beyond it for large and small collision systems.

\noindent\textit{\textbf{Energy correlators of flowing bulk matter:}} 
For a given state $|A\rangle$,
the two-point energy-flow correlator can be decomposed as
~\cite{Cuomo:2025pjp,Barata:2024ukm,Barata:2025fzd,Barata:2026pgh,
Barata:2026aak,Chicherin:2023gxt,Zhao:2025ogc}
\begin{align}
\langle\hat{\mathcal{E}}(\n_1) \, \hat{\mathcal{E}}(\n_2)\rangle=
        \langle\hat{\mathcal{E}}(\n_1)\rangle \, \langle\hat{\mathcal{E}}(\n_2)\rangle+
        \langle\delta\hat{\mathcal{E}}(\n_1) \, \delta\hat{\mathcal{E}}(\n_2)\rangle,
\end{align} 
where we use a shorthand notation $\langle A|{\hat{\mathcal{O}}}|A\rangle=\langle{\hat{\mathcal{O}}}\rangle$ and $\delta \hat{\mathcal{E}}$ denotes the fluctuation of the energy-flow operator around its expectation value in the given many-body state.  The first term on the right-hand side is the disconnected contribution determined by the mean energy flow in the given state, while the second term is  the connected contribution, which encodes correlations beyond the mean energy flow, with different angular regimes probing dynamics ranging from microscopic QFT processes to collective hydrodynamic fluctuations. 

For the case with a finite acceptance window, the single-angle projected EEC is written as
\begin{equation}
  \label{eq: single angle EEC}
    \begin{split}
      \frac{d\Sigma}{d\cos\chi} = &\int_{\mathbf{n}_{1,2}}  \frac{\overline{\left\langle \hat{\mathcal{E}}(\mathbf{n}_1) \hat{\mathcal{E}}(\mathbf{n}_2) \right\rangle}}{\overline{\langle \hat{E}^2\rangle}}  \\
      &\times C(\theta_1)C(\theta_2)\delta(\mathbf{n}_1\cdot \mathbf{n}_2-\cos\chi), 
    \end{split}  
\end{equation}
where $C(\theta)=\Theta(\mu-|\cos\theta|)$ is a pseudorapidity acceptance window imposed by a realistic detector, $\hat{E}=\int_{\bf n}\hat{\mathcal{E}}(\mathbf{n}) C(\theta)$ is the total energy within that region, and the overline denotes the ensemble average running over the different states $|A\rangle$. Since the ensemble average of a product of one-point expectation values is not directly measurable, we reconstruct the disconnected part using the product of the ensemble-averaged energy flows, each obtained through event averaging in the final state:
\begin{align}\label{eq:Sigma disc}
\frac{d\Sigma^{\rm disc}}{d\cos\chi} \equiv & \int_{\n_{1,2}}\frac{\overline{\langle\hat{\mathcal{E}}(\n_1)\rangle}\;\overline{\langle\hat{\mathcal{E}}(\n_2)\rangle}}{\overline{\langle \hat{E}\rangle}^2}\notag\\
&\times C(\theta_1)C(\theta_2)\delta(\n_1\cdot\n_2-\cos\chi)\, ,
\end{align}
where the normalization is chosen such that the disconnected EEC integrates to unity.   
The ensemble-averaged energy flow is independent of the azimuthal angle. The angular structure of the disconnected EEC in Eq.~\eqref{eq:Sigma disc} is therefore determined by the longitudinal profile of the mean energy flow and the pseudorapidity acceptance~\cite{Peng:bulkEECinPreparation}.

In high energy nuclear collisions, the flow harmonics are important soft observables defined through the Fourier expansion of the azimuthal particle distribution~\cite{Schenke:2010rr,Gale:2012rq,Heinz:2013th,Song:2017wtw}.
Analogously, we expand the energy flow in each event in azimuthal Fourier modes, 
\begin{equation}
  \label{eq:Fourier expand energy flow}
  \mathcal{E}(\mathbf{n})=\frac1{2\pi}\left[\mathcal{E}_{0}(\cos\theta)+
                          \sum_{n\not=0}\mathcal{E}_{n}(\cos\theta)e^{-in\phi}\right],
\end{equation}
where the Fourier coefficients satisfy $\mathcal{E}_{n}=\mathcal{E}^*_{-n}$, with their magnitudes related to the energy-weighted flow harmonics through $|\mcal{E}_n|=v_n^E\mcal{E}_0$~\cite{Liu:2015nwa}. 

\begin{figure*}[t] 
\centering
\includegraphics[width=\textwidth]{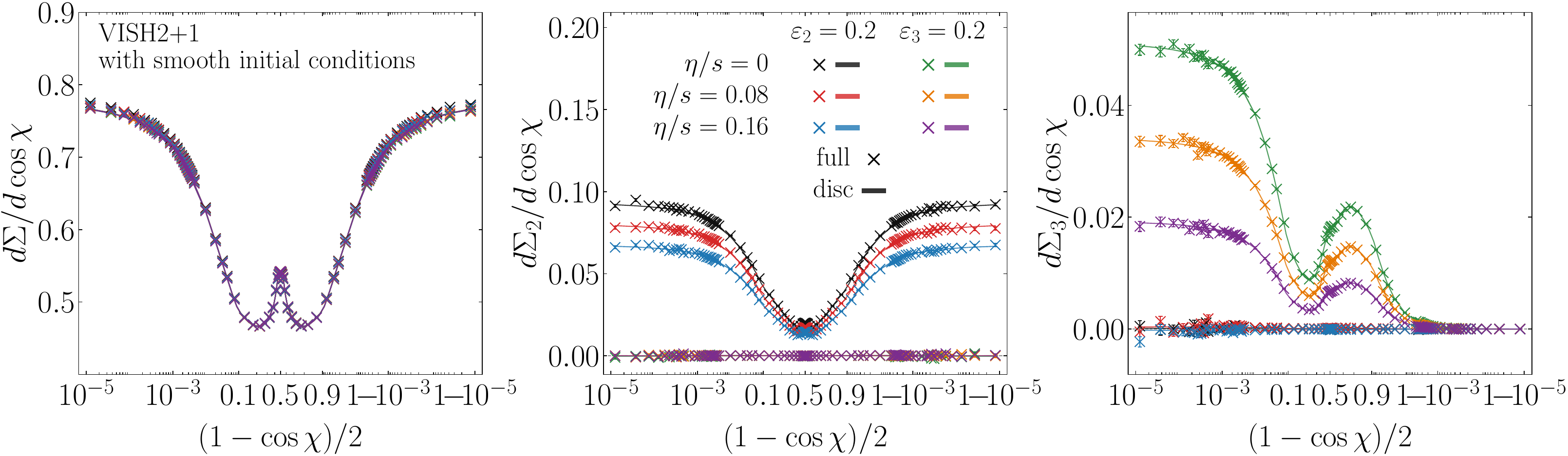}
\caption{Full-angle EEC $d\Sigma/d\cos\chi$ (left),
its elliptic harmonic $d\Sigma_2/d\cos\chi$ (middle), and
triangular harmonic $d\Sigma_3/d\cos\chi$ (right) of soft hadrons with $|\eta|<1$, calculated from viscous hydrodynamics VISH2+1   with smooth initial conditions. The initial eccentricity is set to either $\varepsilon_2=0.2$ and $\varepsilon_3=0$ or $\varepsilon_2=0 $ and $\varepsilon_3=0.2$ and the specific shear viscosity is set to $\eta/s=0,\;0.08$, or $0.16$. Crosses and bands denote the full and disconnected contributions, respectively.}
    \label{fig:smooth Sigman}
\end{figure*}

Following the logic of Refs.~\cite{Barata:2025zku,Barata:2026pgh,Barata:2026aak}, we resolve the azimuthal structure of EECs through EEC harmonics.  To define these, we introduce the energy-weighted event plane angle $\Psi_{n}^E$ giving an azimuthal reference, $\Psi_{n}^E = n^{-1}\mathrm{arg}\left\{ \sum_j E_j e^{in\phi_j}\tilde{C}(\theta_j)\right\}$, 
where the summation runs over particles of one event, and $\tilde{C}(\theta)$ denotes an acceptance region which does not overlap with $C(\theta)$. The EEC harmonics are then defined by
\begin{equation}
    \label{eq:EEC harmonics}
    \begin{split}
        \frac{d\Sigma_n}{d\cos\chi} = & \int_{\mathbf{n}_{1,2}}\sum_{i=1,2}\frac{\overline{{\langle \hat{\mathcal{E}}(\mathbf{n}_1)\hat{\mathcal{E}}(\mathbf{n}_2)\rangle}\cos n \left( \phi_i - \Psi_{n}^E \right)}}{2\overline{\langle \hat{E}^2\rangle}}\\
        &\times C(\theta_1)C(\theta_2)\delta(\mathbf{n}_1\cdot \mathbf{n}_2-\cos\chi)\,,
    \end{split} 
\end{equation}
where the average in the numerator is taken over events.

The corresponding disconnected harmonics can be constructed from the event-averaged Fourier coefficients of mean energy flow $\overline{\mcal{E}_n^{\mathrm{ref}}}\equiv \mathrm{Re}\;\overline{\mathcal{E}_ne^{-in\Psi_n^E}}$. In our calculations, we find that retaining terms linear in the Fourier coefficient $\mcal{E}_n$ is sufficient to describe the disconnected EEC harmonics~\cite{Peng:bulkEECinPreparation}:
\begin{equation}
  \label{eq:factorized EEC harmonics}
   \begin{split}
    \frac{d\Sigma_n^{\rm disc}}{d\cos\chi} \approx& \int_{-\mu}^{\mu} d\mu_1 d\mu_2 \int_0^{2\pi}\frac{d\Delta\phi}{2\pi}\,2\cos^2 \frac{n\Delta\phi}2\\
    &\times \, \frac{\overline{\mcal{E}_{0}(\mu_1)}\,\overline{\mcal{E}^{\mathrm{ref}}_{n}(\mu_2)}}{\overline{\langle \hat{E}\rangle}^{\,2}} \delta(\mathbf{n}_1\cdot \mathbf{n}_2-\cos\chi),
   \end{split}
\end{equation}
where $\mu_{1,2}=\cos\theta_{1,2}$ and $\Delta\phi = \phi_1-\phi_2$. We apply the same normalization as the disconnected expression of single-angle EEC. When the disconnected contribution dominates, the EEC harmonics are governed by the anisotropies of the mean energy flow and are closely related to the energy-weighted flow harmonics. The connected contribution, by contrast, retains sensitivity to correlations that are not encoded in the mean energy flow.

\noindent\textit{\textbf{Models for flow and non-flow systems:}} 
In this paper, we implement the VISH2+1 viscous hydrodynamic model, the iEBE-VISHNU hybrid model and PYTHIA to generate final hadrons and study the soft EEC and its harmonics of final-state hadrons\footnote{A comparison between the (2+1)-dimensional VISH2+1 and (3+1)-dimensional MUSIC calculations is presented in the supplemental material to show that the longitudinal structure of the evolving fluid does not alter the main conclusions of this paper.}. In more detail,
VISH2+1~\cite{Heinz:2005bw,Song:2007ux,Song:2007fn} is a (2+1)-dimensional viscous hydrodynamic model with longitudinal boost invariance to simulate the expansion of the QGP fluid. iEBE-VISHNU~\cite{Song:2010aq,Shen:2014vra} couples VISH2+1 to UrQMD~\cite{Bass:1998ca,Bleicher:1999xi} at a switching temperature (around $T_{\mathrm{sw}}=150 \mathrm{MeV}$) for subsequent hadronic scatterings and decays. With the fluid expansion, VISH2+1 and the iEBE-VISHNU generate collective flow in the final-state hadron distributions.  For comparison, we run PYTHIA~8.3~\cite{Bierlich:2022pfr} with its Angantyr extension for nuclear collisions~\cite{Bierlich:2018xfw}. This model includes multiparton interactions, parton showers, color reconnection, string fragmentation, decays of unstable hadrons, the correlations of which are basically non-flow.

To isolate the response to the initial geometry and shear viscosity, we first perform VISH2+1 simulations with smooth initial conditions, in which the initial energy density has a Gaussian transverse profile with an elliptic or triangular deformation, tuned such that the initial eccentricities satisfy either $\varepsilon_2=0.2$ and $\varepsilon_3=0$
or $\varepsilon_2=0$ and $\varepsilon_3=0.2$~\cite{Peng:bulkEECinPreparation}. 
The fluid systems freeze out into hadrons at $T_{\mathrm{frz}}=150 \mathrm{MeV}$ without hadronic scatterings and decays. 
For more realistic simulations of collision systems of different sizes, we use iEBE-VISHNU with fluctuating initial conditions generated by the T$_\mathrm{R}$ENTo model \cite{Moreland:2014oya} for Pb+Pb collisions at $\sqrt{s_{NN}}=5.02$ TeV, O+O collisions at $\sqrt{s_{NN}}=5.36$ TeV, and p+p collisions at $\sqrt{s}=5.02$ TeV. The parameters of T$_\mathrm{R}$ENTo and VISH2+1 are taken from the Bayesian calibration to soft-hadron data in Pb+Pb and p+Pb collisions at $5.02~\mathrm{TeV}$~\cite{Moreland:2018gsh}. Correspondingly, PYTHIA/Angantyr is run with the default tune~\cite{Skands:2014pea} for p+p, O+O, and Pb+Pb collisions.

\begin{figure*}[t] 
    \centering
    \includegraphics[width=\textwidth]{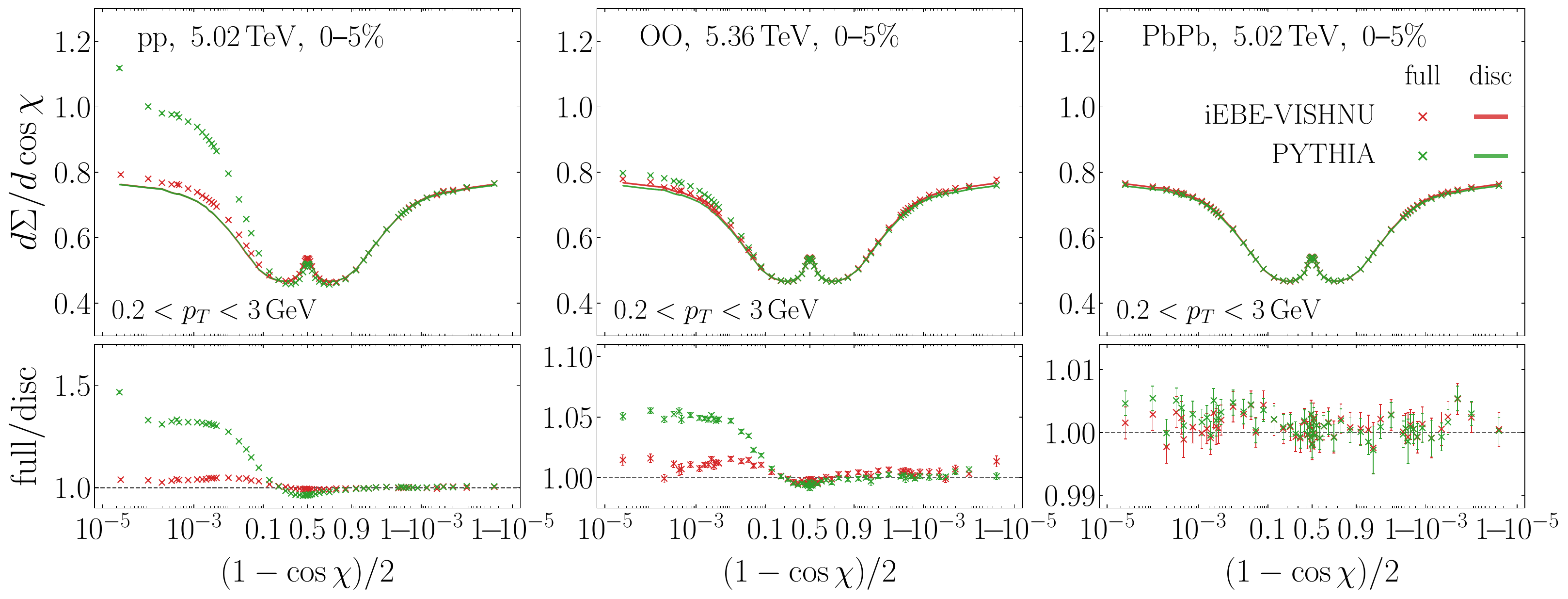}
    \caption{Full and disconnected EECs for soft hadrons with $|\eta|<1$ and a transverse momentum cut $0.2<p_T<3~\mathrm{GeV}$ in p+p, O+O and Pb+Pb collisions, calculated using PYTHIA and iEBE-VISHNU. Crosses and bands denote the full and disconnected contributions, respectively. The lower panels show their ratio.}
    \label{fig:system-size}
\end{figure*}

\noindent\textit{\textbf{Soft EECs across collision systems:}}
We now study the soft EEC of final-state hadrons for large and small systems at the LHC. With the final hadrons produced in experiment or generated by a model, the full EEC and its harmonics from Eqs.~\eqref{eq: single angle EEC} and \eqref{eq:EEC harmonics} can be evaluated as
\begin{align} \label{eq:hadronic}
    \frac{d\Sigma}{d\cos\chi}&=\frac{1}{\sum\limits_{\text{events}} \sum_{i<j}E_iE_j}\,
                               \sum\limits_{\text{events}}\sum_{i<j} E_iE_j\,\notag\\
    &\hspace{0.4cm}\times C(\theta_i)C(\theta_j)\,\delta(\mathbf n_i\cdot\mathbf n_j-\cos\chi),\\
    \frac{d\Sigma_n}{d\cos\chi} &=\frac{1}{\sum\limits_{\text{events}} \sum_{i<j}E_iE_j}\,\sum\limits_{\text{events}}
                                  \sum_{i<j}E_iE_j\,\notag\\
    &\hspace{0.4cm}\times C(\theta_i)C(\theta_j)\,\delta(\mathbf n_i\cdot\mathbf n_j-\cos\chi)\notag\\
    &\hspace{0.4cm}\times\frac12 \left[\cos n(\phi_i-\Psi_n^E) + \cos n(\phi_j-\Psi_n^E) \right]
\end{align}
where {the sum $\sum_{i<j}$} runs over distinct charged particle pairs in each event.  {The event-averaged Fourier coefficients $\overline{\mcal{E}_n^{\mathrm{ref}}}$ are calculated by}
\begin{align}\label{eq:hadronic Enref}
    \overline{\mcal{E}_n^{\rm{ref}}}(\cos\theta)=&\frac{1}{N_{\rm{ev}}} \sum_{\rm{events}}\sum_j  C(\theta_j) E_j \cos n(\phi_j-\Psi_n^E)\notag\\
    &\hspace{1cm}\times \delta(\cos\theta_j-\cos\theta),
\end{align}
with $\overline{\mcal{E}_0^{\mathrm{ref}}}=\overline{\mcal{E}_0 }$ and $N_{\mathrm{ev}}$ representing the number of events. The disconnected contributions to single-angle EEC and its harmonics are then calculated by substituting Eq.~\eqref{eq:hadronic Enref} into Eq.~\eqref{eq:Sigma disc} and Eq.~\eqref{eq:factorized EEC harmonics}, respectively.

For the evolving QCD matter, we first check the sensitivity of soft EEC 
to the initial-state geometry and to the specific
shear viscosity.   Figure~\ref{fig:smooth Sigman} 
shows the EEC and its elliptic and
triangular harmonics from VISH2+1 hydrodynamics with smooth initial
conditions. In all three panels, the disconnected results agree with the full 
results within uncertainties, since final-state hadrons in the pure hydrodynamic calculation are independently described by single-particle distributions.
The angular structure of the single-angle EEC is governed primarily by
the longitudinal mean energy flow within a finite pseudorapidity acceptance, which 
is insensitive to both $\varepsilon_n$ and $\eta/s$.
The peaks near the two endpoints and the small bump near $\chi=\pi/2$ can be qualitatively
described by an azimuthally symmetric and boost-invariant profile,
whose mean energy flow is proportional to $1/\sin^3\theta$~\cite{Peng:bulkEECinPreparation}.
In contrast, the harmonics of soft EEC behave as traditional energy-weighted flow harmonics with their magnitude decreasing with increasing $\eta/s$. Meanwhile, $d\Sigma_2/d\cos\chi$ and $d\Sigma_3/d\cos\chi$ also show a strong response to the corresponding initial eccentricities
$\varepsilon_2$ and $\varepsilon_3$, respectively.

We next turn to realistic fluctuating initial conditions and use event-by-event simulations from iEBE-VISHNU and PYTHIA to examine how the relative importance of disconnected and connected contributions to the EEC evolves from large to small collision systems.
Figure~\ref{fig:system-size} plots
the full and disconnected EECs in p+p, O+O, and Pb+Pb
collisions. For iEBE-VISHNU, the disconnected results almost overlap with the full ones over the whole angular range in all three systems, similar to the results from hydrodynamics with smooth initial conditions in Fig.~2. Only small deviations ($<5\%$) appear at small angles in p+p collisions, which arise from the limited correlations induced by hadronic rescattering and decays in iEBE-VISHNU. For PYTHIA, the full and disconnected results agree at large angles in all three systems, while the small-angle deviation increases as the collision system becomes smaller. This trend reflects the dilution of connected correlations by the growing number of uncorrelated pairs as system size and multiplicity increase. The substantially larger deviations between full and disconnected results at small angles in small collision systems in PYTHIA than in iEBE-VISHNU  indicate the effects from additional correlations, such as mini-jets, string fragmentation, and resonance decays.

\begin{figure*}[t] 
    \centering
    \includegraphics[width=\textwidth]{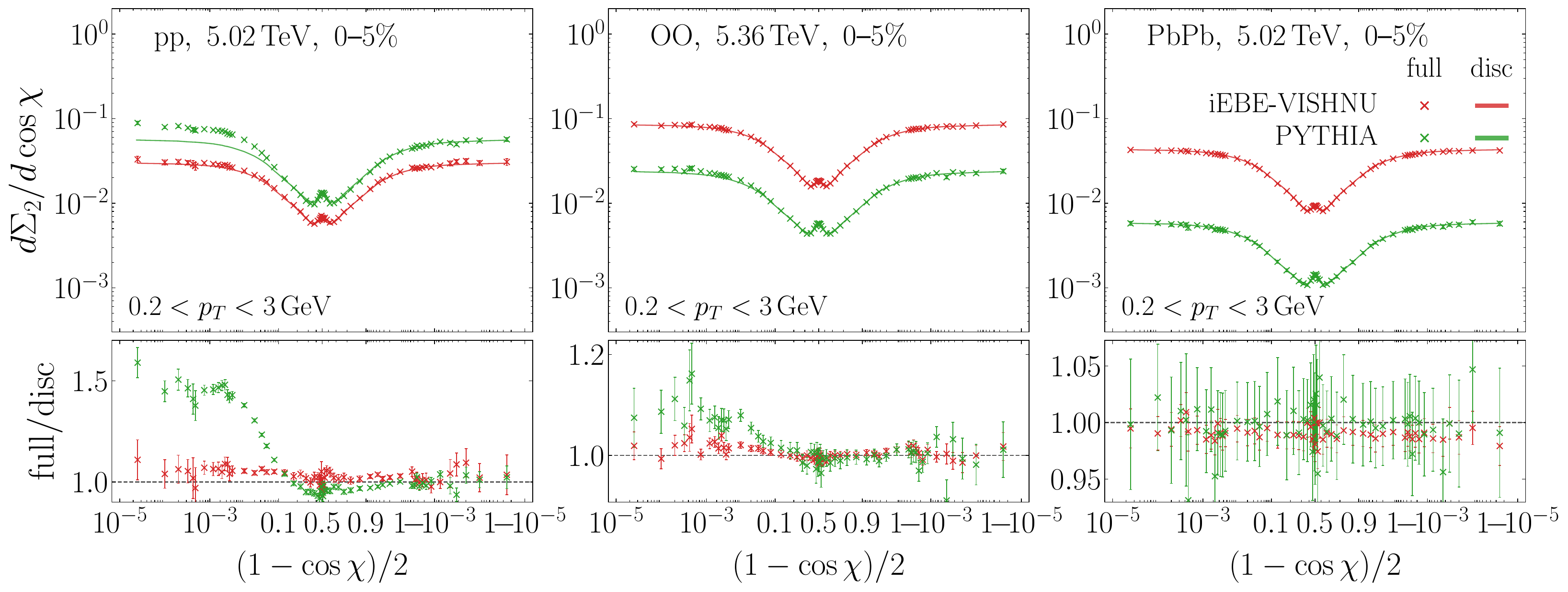}
    \caption{Full and disconnected elliptic EEC harmonics $d\Sigma_2/d\cos\chi$ for soft hadrons with $|\eta|<1$ and a transverse momentum cut $0.2<p_T<3~\mathrm{GeV}$ in p+p, O+O, and Pb+Pb collisions, calculated using iEBE-VISHNU and PYTHIA. Crosses and bands denote the full and disconnected contributions, respectively. The lower panels show their ratio.}
    \label{fig:sigma2-across-systems}
\end{figure*}

The same comparison can then be made for the EEC harmonics. Figure~\ref{fig:sigma2-across-systems} shows the full and disconnected elliptic harmonics $d\Sigma_2/d\cos\chi$ in  p+p, O+O, and Pb+Pb collisions, calculated from iEBE-VISHNU and PYTHIA. For iEBE-VISHNU, the magnitude of
$d\Sigma_2/d\cos\chi$ is comparable for all three systems, reflecting the development of elliptic mean energy flow through collective expansion. The disconnected results 
again almost overlap with the full ones over the 
whole angular range, with only a few percent deviations
at small angles in p+p collisions as in Fig.~\ref{fig:system-size}. 
In PYTHIA, the magnitude of the elliptic EEC harmonic drastically decreases from p+p to O+O and Pb+Pb, which reflects its nonflow origin, with correlations increasingly diluted as the multiplicity grows in the absence of collective expansion.
The disconnected contribution still describes the full harmonic at large angular
separations in all systems as in Fig.~\ref{fig:system-size}, while the small-angle deviations are larger than the ones in Fig.~\ref{fig:system-size}, especially for 
p+p collisions. The EEC and its harmonics can therefore probe collective flow from expanding fluid systems and reveal correlations beyond flow, especially for small collision systems.

\noindent\textit{\textbf{Conclusion:}} In this \textit{Letter}, we propose energy-energy correlators (EECs) of soft hadrons and their azimuthal harmonics as new probes for collective flow and correlations beyond flow for large and small systems created in high energy nuclear collisions. Built on the field-theoretic definition of the energy flow, the soft EEC separates into a disconnected contribution determined by the mean energy flow and a connected contribution encoding correlations beyond it. For fluid systems described by viscous hydrodynamics and the hybrid model iEBE-VISHNU, the EEC and its harmonics are dominated by the disconnected
contribution. The single-angle EEC is governed primarily by the longitudinal
mean energy-flow profile, whereas the EEC harmonics show clear sensitivity
to the shear viscosity and the initial-state geometry, as do the traditional
flow harmonics. In PYTHIA, by contrast, the disconnected contribution
dominates at large angular separations, with connected correlations becoming
increasingly important at small angles from Pb+Pb to O+O and p+p collisions.
For the elliptic EEC harmonics, the PYTHIA results drop by about an order of 
magnitude from p+p to Pb+Pb collisions due to the
dilution of non-flow correlations, whereas the iEBE-VISHNU results remain
comparable across systems. Meanwhile, small-angle deviations in p+p collisions remain more pronounced in PYTHIA than in iEBE-VISHNU. This comparison demonstrates
that the EEC and its harmonics are new soft observables to distinguish
collective flow from non-flow correlations, particularly in small collision
systems.

In experiment, the soft EECs and the azimuthal harmonics could be measured in heavy ion and light ion collisions at RHIC and the LHC, which could help probe the collectivity and the QGP formation in large and small collision systems. More generally, they could also serve as new observables for studying fluctuations and correlations in the initial state, bulk evolution and the transport coefficients of the QCD matter, as well as the medium response to jets.  Further studies could identify the angular scales at which different physical mechanisms become important and clarify the origins of the connected contribution. The same framework could extend to three-point and higher-point energy correlators, allowing the angular and shape dependence of multiparticle correlations to be investigated. These developments would establish the soft energy correlators as a bridge between the microscopic dynamics of QCD and the macroscopic collectivity of the QCD matter produced in high energy nuclear collisions.

\noindent\textit{\textbf{Acknowledgment:}} We thank G. Giacalone, A. Grassi, C. Iossa, M. Kuzmin, J. M. Silva, L. Tizzano, and A. Zhiboedov for helpful discussions. JB was supported by the U.S. Department of Energy under Contract No. DE-SC0012704. The work of YP and HS is supported by the NSFC under grants No.12575138 and No.12247107. The work of AVS is supported by the Spanish Ministry of Science, Innovation and Universities through grant PID2025-174203NB-I00 (MICIU/AEI/10.13039/501100011033), by Fundação para a Ciência e a Tecnologia (FCT) through grants 2022.06565.CEECIND and 2023.15319.PEX, and by the Basque Government through grant IT1977-26. AVS also acknowledges support from Ikerbasque, Basque Foundation for Science. This work was performed at the Tianhe Supercomputer Center in Tianjin (Tianhe New Generation Supercomputer), and the High-Performance Computing Platform of Peking University.

\bibliographystyle{bibstyle.bst}
\bibliography{EEC_ref.bib,hydro_ref.bib}

\clearpage
\onecolumngrid

\section*{\protect\Large Supplemental material}

\begin{figure*}[h]
    \centering
    \includegraphics[width=\textwidth]{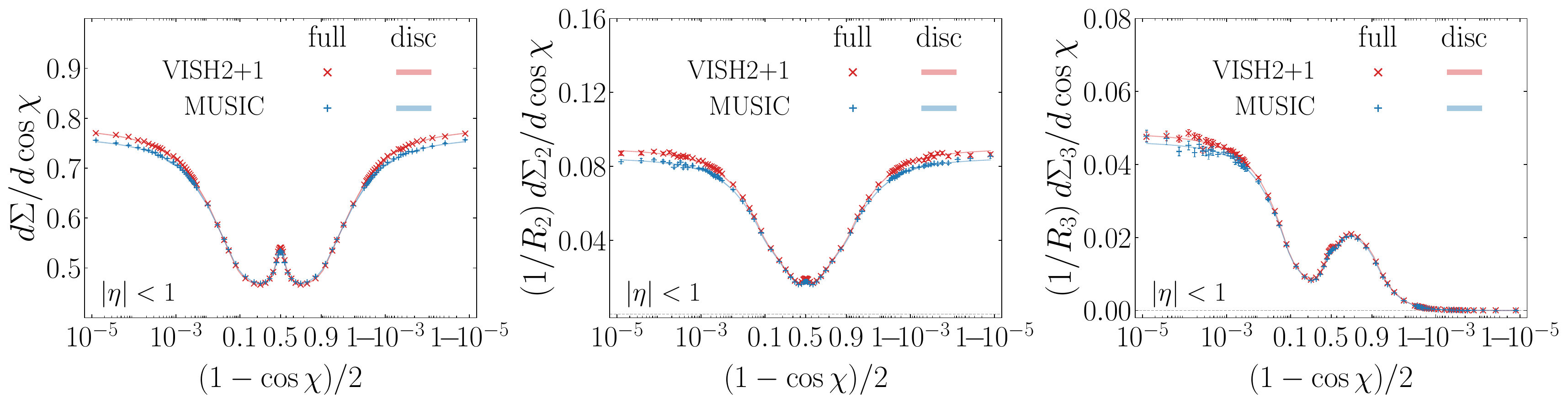}
    \caption{Full-angle EEC $d\Sigma/d\cos\chi$ (left), and the resolution-corrected elliptic harmonic $(1/R_2)\,d\Sigma_2/d\cos\chi$ (middle) and triangular harmonic $(1/R_3)\,d\Sigma_3/d\cos\chi$ (right) of soft hadrons with $|\eta|<1$, calculated from VISH2+1 and MUSIC with smooth initial conditions. Crosses and bands denote the full and disconnected contributions, respectively. 
    }
    \label{fig:vishnu-music-pbpb}
\end{figure*}

\subsection{Effects of longitudinal expansion}
In this subsection, we compare (2+1)-dimensional VISH2+1 and (3+1)-dimensional MUSIC hydrodynamic calculations to evaluate the effects of longitudinal expansion on the soft EEC and its harmonics. For a controlled comparison, both calculations use $\eta/s=0.08$ and a freeze-out temperature of $150~\mathrm{MeV}$. The transverse energy-density profile in MUSIC at $\eta_s=0$ is chosen to match that in VISH2+1. The longitudinal profile of the initial energy density has a flat central plateau for $|\eta_s|<1.5$ and Gaussian falloffs at larger $|\eta_s|$. The initial eccentricities are set to ($\varepsilon_2=0.2,\;\varepsilon_3=0$) for the elliptic harmonic and ($\varepsilon_2=0,\;\varepsilon_3=0.2$) for the triangular harmonic.

Figure~\ref{fig:vishnu-music-pbpb} compares the full-angle EEC and its elliptic and triangular harmonics from VISH2+1 and MUSIC, for soft hadrons with $|\eta|<1$ and with the event-plane angles $\Psi_n^E$ reconstructed from hadrons within $1<|\eta|<2$. As in the event-plane method for traditional flow harmonics~\cite{CMS:2012zex}, the finite event-plane resolution suppresses the EEC harmonics by an overall factor. This leads to the suppression of the MUSIC harmonics with overall magnitudes about $10\%$ (n=2) and $20\%$ (n=3) smaller than those from VISH2+1. To account for this effect, we calculate the resolution-corrected EEC harmonics with the resolution evaluated with three separate pseudorapidity regions~\cite{CMS:2012zex}:
\begin{equation} \label{eq:ep-resolution}
    R_n = \left[\frac{\left\langle\cos n(\Psi_n^{E,A}-\Psi_n^{E,B})\right\rangle_{\mathrm{ev}}
          \left\langle\cos n(\Psi_n^{E,A}-\Psi_n^{E,C})\right\rangle_{\mathrm{ev}}}
         {\left\langle\cos n(\Psi_n^{E,B}-\Psi_n^{E,C})\right\rangle_{\mathrm{ev}}}\right]^{1/2}.
    \end{equation}
Here $\Psi_n^E$ in Eq.(9) with $1<|\eta|<2$ is denoted by $\Psi_n^{E,A}$. $\Psi_n^{E,B}$ and $\Psi_n^{E,C}$ are the energy-weighted event-plane angles calculated within $-1<\eta<0$ and $0<\eta<1$, respectively, and the brackets denote averages over events.

Figure~\ref{fig:vishnu-music-pbpb} shows that in all three panels, the disconnected results agree with the full results within uncertainties for both VISH2+1 and MUSIC, as in Fig.~\ref{fig:smooth Sigman}, so the dominance of the disconnected contribution does not rely on boost invariance in the fluid system. For the full-angle EEC (left panel), the MUSIC results almost follow the VISH2+1 result over the whole angular range with about $2\%$ difference, reflecting mild effects from the longitudinal evolution on the soft EEC at mid-rapidity.
For the EEC harmonics with resolution correction (middle and right panels), the differences between the MUSIC and VISH2+1 results are within $5\%$. These differences are much smaller than the differences between the PYTHIA and iEBE-VISHNU results in Fig.~4, so the longitudinal structure of the evolving fluid does not change the main conclusions of this Letter.

\clearpage

\begin{figure*}[ht]
    \centering
    \includegraphics[width=\textwidth]{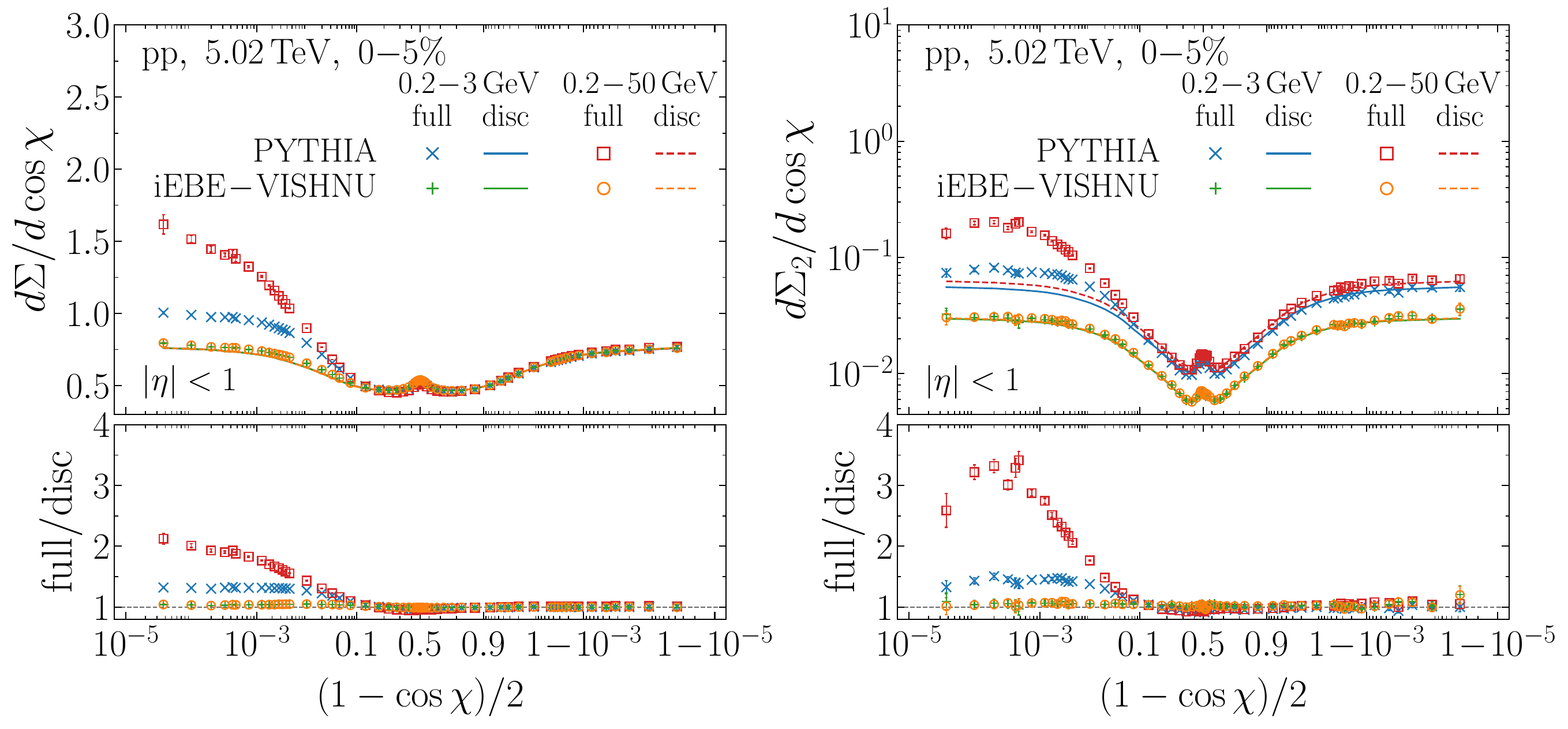}
    \caption{Full and disconnected EECs $d\Sigma/d\cos\chi$ (left) and elliptic EEC harmonics $d\Sigma_2/d\cos\chi$ (right) for charged hadrons with different transverse momentum cuts $0.2<p_T<3~\mathrm{GeV}$ and $0.2<p_T<50~\mathrm{GeV}$, calculated from PYTHIA and iEBE-VISHNU. Symbols and lines denote the full and disconnected contributions, respectively. The lower panels show their ratio.}
    \label{fig:pp-pt-range}
\end{figure*}

\subsection{Dependence on the transverse-momentum range}
\label{sec:pt-range-dependence}

In this subsection, we check how the soft EEC and its harmonics depend on the transverse momentum cut of the charged hadrons. In Figs.~\ref{fig:system-size} and~\ref{fig:sigma2-across-systems}, we use the same transverse momentum cut $0.2<p_T<3~\mathrm{GeV}$ for a consistent comparison between PYTHIA and iEBE-VISHNU, and this range covers almost all the hadrons produced in hydrodynamic systems. In PYTHIA, however, perturbative multiparton interactions do not restrict the scattered partons to low $p_T$. Hadrons produced at higher transverse momenta may give sizable contributions to the EEC and its harmonics. To assess this effect, we perform comparison calculations for the default range $0.2<p_T<3~\mathrm{GeV}$ and an extended range $0.2<p_T<50~\mathrm{GeV}$, while keeping all other settings unchanged.

Figure~\ref{fig:pp-pt-range} shows the full and disconnected EECs and the elliptic EEC harmonics in $0$--$5\%$ central p+p collisions at $\sqrt{s}=5.02~\mathrm{TeV}$, calculated from iEBE-VISHNU and PYTHIA with these two transverse momentum ranges. For iEBE-VISHNU, the full and disconnected results are almost unchanged when the range is extended, since the hydrodynamic systems produce very few hadrons above $3~\mathrm{GeV}$. In PYTHIA, by contrast, including higher-$p_T$ hadrons strongly modifies the small-angle behavior of the full results. For the single-angle EEC (left panel), the full $d\Sigma/d\cos\chi$ is enhanced at small angles, while the disconnected results change little. Such an enhancement with the extended $p_T$ range for the full results thus reflects the connected correlations among hadrons from mini-jets, etc. For the EEC harmonics (right panel), the overall magnitude of the disconnected $d\Sigma_2/d\cos\chi$ (solid and dashed lines) from PYTHIA increases moderately at both small and large angles with the increase of the transverse momentum range, due to the inclusion of high-$p_T$ hadrons with larger elliptic harmonics. At small angles, the full $d\Sigma_2/d\cos\chi$ increases substantially, with a relative enhancement considerably larger than that of $d\Sigma/d\cos\chi$. This indicates that the EEC elliptic harmonic at small angles is more sensitive to small-angle correlations involving high-$p_T$ hadrons from the non-flow systems.

\end{document}